\documentclass[review]{elsarticle}

\journal{Nuclear Physics A}

\begin{document}

\begin{frontmatter}

\title{Coulomb corrections to Fermi beta decay in nuclei}

\author{Naftali Auerbach}
\address{School of Physics and Astronomy, Tel Aviv University, Tel Aviv 69978, Israel}
\ead{auerbach@tauex.tau.ac.il}
\author{Bui Minh Loc\fnref{myfootnote}\corref{mycorrespondingauthor}}
\address{Division of Nuclear Physics, Advanced Institute of Materials Science, Ton Duc Thang University, Ho Chi Minh City, Vietnam \\
Faculty of Applied Sciences, Ton Duc Thang University, Ho Chi Minh City, Vietnam}
\fntext[myfootnote]{Present address: Center for Exotic Nuclear Studies, Institute for Basic Science (IBS), Daejeon 34126, Korea}
\cortext[mycorrespondingauthor]{Corresponding author}
\ead{buiminhloc@tdtu.edu.vn}

\begin{abstract}
We study the influence of the Coulomb force on the Fermi beta-decays in nuclei. This work is composed of two main parts. In the first part, we calculate the Coulomb corrections to super-allowed beta decay. We use the notion of the isovector monopole state and the self-consistent charge-exchange Random Phase Approximation to compute the correction. In the second part of this work, we examine the influence of the anti-analog state on isospin mixing in the isobaric analog state and the correction to the beta-decay Fermi transition.
\end{abstract}

\begin{keyword}
super-allowed beta decay\sep isovector monopole state\sep anti-analog state \sep isospin mixing
\end{keyword}

\end{frontmatter}

%\linenumbers

\section{Introduction}
In a number of studies attempts are made to determine corrections one has to introduce in the evaluation of the beta-decay matrix elements. In particular super-allowed transitions in $T = 1, T_z = +1$ (or $T_z = -1$) nuclei \cite{TH, Auerbach09, Liang09, Satula11, Rodin13, Xayavong18, Ormand95, HT20} are extensively studied theoretically and experimentally. These corrections are important because using the measured $ft$ values one can relate these to the $u$-quark to $d$-quark transition matrix element $V_{ud}$ in the Cabibbo-Kobayashi-Maskawa (CKM) matrix. In the Standard Model this matrix fulfils the unitarity condition, the sum of squares of the matrix elements in each row (column) is equal to one, as for example:
\begin{equation}
 V_{ud}^2 + V_{us}^2 + V_{ub}^2 = 1.
\end{equation}

In order to determine the $V_{ud}$ term using the experimental $ft$ values in the super-allowed beta-decay, one must introduce a number of corrections \cite{TH}. In this paper, similarly to reference \cite{Auerbach09} we will consider one important aspect of it, namely the Coulomb correction. There have been a number of works that dealt with this problem using different methods \cite{TH, Auerbach09, Liang09, Satula11, Rodin13, Xayavong18, Ormand95, HT20} and more. In particular, the authors of \cite{TH, HT20} have devoted a considerable amount of work to study the influence of the Coulomb interaction on the $ft$ values. Of course in any of these studies, there are some approximations involved. One of the main issues is the way the Coulomb force introduces admixtures of higher excitations into the parent and its analog state. This was the main topic of reference \cite{Auerbach09}. The Coulomb force admixes particle-hole ($ph$) states, mostly of $2\hbar \omega$ at unperturbed energy positions. There is, however, a $ph$ interaction that changes the situation, creating a collective state. In the case of a one-body Coulomb potential, the excitation caused by it leads to $J=0^+, T = 1$ $ph$ states. In the isovector channel, the $ph$ interaction is repulsive, and therefore there is an upward energy shift. The resulting collective state is the isovector monopole (IVM) giant resonance. The shift is substantial as many Random Phase Approximation (RPA) studies indicate \cite{AuerbachKlein83, Auerbach83, Loc19} about $1\hbar \omega$, and in some other studies even higher \cite{BMVolI, Auerbach72}, $2\hbar \omega$ shifts. In the RPA, the energy weighted sum rule is conserved, and therefore the upward energy shift will reduce the strength. The amount of Coulomb mixing is determined by the strength divided by the energy squared. Therefore in a hand waving argument this amount will be reduced by a factor $(2/3)^3 = 8/27$ in the RPA, compared to the calculation in which unperturbed $2\hbar \omega$ $ph$ excitations are used. As we will see in the next sections the actual calculations confirm this rough estimate. There are additional drawbacks in the shell model approaches \cite{TH} as pointed out in reference \cite{Miller08} which were avoided in \cite{Auerbach09}. We will now briefly outline the main steps in the theory given in \cite{Auerbach09}.

\section{Coulomb Mixing and the isovector monopole}
 The Fermi beta decay matrix element between the ground state and its isobaric analog state (IAS) we write in the form:
 \begin{equation}\label{eq1}
 |M_F|^2 = |M_F^0|^2(1 - \delta_C)
 \end{equation}
where $M_F$ is the physical Fermi matrix element:
\begin{equation}\label{me}
M_F = \langle \Psi_1 |T_+ | \Psi_2 \rangle.
\end{equation}
$| \Psi_1 \rangle$ and $| \Psi_2 \rangle$ are the parent and daughter physical states. The symbol $M_F^0$ stands for the Fermi matrix element obtained in the limit when in the Hamiltonian all the charge-dependent parts are put to zero, and the wave functions are eigenstates of the charge-independent Hamiltonian $H_0$. The symbol $\delta_C$ is the Coulomb correction. The eigenstates of this Hamiltonian with isospin $T$ and $T_z$ will be denoted as $| T, T_z\rangle$ and:
\begin{equation}
 H_0 | T, T_z \rangle = E_T | T, T_z \rangle.
\end{equation}
The $2T+1$, components with different $T_z$ values are degenerate, the action of the isospin lowering and raising operators, $T_{-}$, $T_+$ gives:
\begin{eqnarray}
 T_{-} | T, T \rangle = \sqrt{2T} |T, T - 1 \rangle, \nonumber \\
 T_{+} | T, T-1 \rangle = \sqrt{2T} |T, T \rangle.
\end{eqnarray}
In this case $M_F^0 = \sqrt{2T}$. We introduce now a charge-dependent part $V_{\rm{CD}}$. The dominant part of the charge-dependent interaction is the charge asymmetric one-body Coulomb potential $V_C$ (While the charge-dependent components of the two-body nuclear force might be important in changing the relative spacing of levels in the analog nucleus its influence in isospin mixing in the ground state or IAS is expected to be small).

The one-body Coulomb potential will now admix into the ground state and its IAS the IVM \cite{Auerbach09, Auerbach83}. In perturbation theory the effect of the charge-dependent part on the wave functions of the two members of the isospin multiplet, $|T, T \rangle$ and $| T, T-1 \rangle$ will be:
\begin{eqnarray}
 \Psi_1 &=& N_1^{-1}(|T, T \rangle + \varepsilon_T| M_{T,T}\rangle + \varepsilon_{T+1} |M_{T+1,T}\rangle), \label{psi1}\\
 \Psi_2 &=& N_2^{-1}(|T, T-1 \rangle + \eta_{T-1}|M_{T-1, T-1} \rangle \nonumber \\ && + \eta_{T}|M_{T,T-1} \rangle + \eta_{T+1} |M_{T+1,T-1} \rangle), \label{psi2}
\end{eqnarray}
where $|M_{T',T'_z} \rangle$, are the $T', T'_z$ components of the IVM, and where
\begin{equation}
 N_1 = \sqrt{1 + \varepsilon_T^2 + \varepsilon_{T+1}^2},
\end{equation}
and
\begin{equation}
 N_2 = \sqrt{1 + \eta_{T-1}^2 + \eta_T^2 + \eta_{T+1}^2},
\end{equation}
with
\begin{equation}
 \varepsilon_i = \frac{\langle T,T|V_C|M_{T+i,T}\rangle} {E_{M_{T+i, T}} - E_0}, \quad i = 0, 1,
\end{equation}
where $E_0$ is the g.s. energy in this nucleus,
\begin{equation}
 \eta_i = \frac{\langle T, T-1 |V_C|M_{T+i,T-1}\rangle} {E_{M_{T+i, T-1}} - E_1}, \quad i = -1, 0, 1.
\end{equation}
Here $E_1$ is the energy of the analog state. One can write these as:
\begin{eqnarray}
 \varepsilon_i &=& \langle T, T, 1, 0| T+i, T \rangle \frac{\langle T+i|| V_C||T \rangle}{E_{M_{T+i,T}} - E_0}, \label{epsilon} \\
 \eta_i &=& \langle T, T, 1, 0| T+i, T-1 \rangle \frac{\langle T+i|| V_C ||T \rangle}{E_{M_{T+i,T-1}} - E_1}. \label{eta}
\end{eqnarray}
Introducing the Clebsch-Gordan coefficients and assuming that the reduced matrix elements are equal, one arrives (\ref{eq1}) at the simple expression: 
\begin{equation}\label{me2}
 \langle \Psi_1 | T_+| \Psi_2 \rangle^2 = 2T \left[1 - 4(T+1) \frac{V_1}{\xi \hbar \omega A} \varepsilon_1^2 \right]^2
\end{equation}
and
\begin{equation}\label{deltacend}
\delta_C = 8(T+1) \frac{V_1}{\xi \hbar\omega A} \varepsilon_1^2.
\end{equation}
Here $\xi\hbar\omega$ is the energy of the IVM in the parent nucleus, $V_1$ is the symmetry energy parameter determined from the equation 
\begin{equation}\label{Vsym}
 E_{M_{T+1, T}} - E_{M_{T,T}} \approx V_1 \frac{N-Z}{A},
\end{equation}
and $\varepsilon_1^2$ is the admixture of the $T+1$ component of the IVM in the parent nucleus. We should emphasize that the result in eq.~(\ref{me2}) is dependent implicitly on all the various admixture in eq.~(\ref{psi1}, \ref{psi2}) and (\ref{epsilon}, \ref{eta}).

The assumption of equal reduced matrix elements in deriving eq.~(\ref{me2}) is approximate. The differences between the reduced matrix elements for different isospin components increase with the increasing number of excess neutrons. See \cite{Auerbach83, Loc19} and references therein. For nuclei with low neutron excess, in particular, for super-allowed decays $(N-Z=2)$, this is a very good approximation. We apply here eq.~(\ref{me2}, \ref{deltacend}) to super-allowed transitions only.

\section{Results of the Coulomb corrections to super-allowed beta decay $\delta_C$}
In reference, \cite{Auerbach09}, the calculations of $\delta_C$ were based on values of isospin mixing derived from some general sum rules and not on detailed microscopic computations of isospin impurities. One calculation presented there has relied on a schematic microscopic model \cite{Yeverechyahu76} which was introduced in the 1970s. We return to the subject of Coulomb corrections because, at present new, more advanced methods to calculate isospin mixing in low-lying nuclear states are available. We mainly rely on the recently published article \cite{Loc19} about isospin impurities calculated using microscopic theories and new types of interactions. Using the formalism described in the previous section we apply equations (\ref{me2}, \ref{deltacend}) to compute the values of $\delta_C$ for a number of nuclei through the periodic table. We concentrate on super-allowed beta-decay transitions. The calculations are performed using the Hartree-Fock (HF) RPA. For open-shell nuclei, one should take into account the pairing correlations and so one has to use the Quasi-particle Random Phase Approximation (QRPA). However in the case of only two nucleons outside the closed shells, one can limit ourselves to RPA. So for the super-allowed transitions (in $T=1$ nuclei), we proceed with our calculation the following way. We calculate in the charge-exchange HF-RPA \cite{Loc19}, the distribution of the IVM strength in the $N=Z$ closed-shell nuclei $^{40}_{20}$Ca, $^{56}_{28}$Ni, $^{80}_{40}$Zr, and $^{100}_{\phantom{1}50}$Sn. In these cases, the IVM has only a $T=1$ isospin. We compute the Coulomb mixing of the IVM into the ground states of these nuclei, (see for details reference \cite{Loc19}) and denote the amount of isospin admixture as $\bar{\varepsilon}^2$. The results are presented in Table \ref{tab1}. 

In the neighboring $T=1$ nuclei, $^{42}_{20}$Ca, $^{58}_{28}$Ni, $^{82}_{40}$Zr, and $^{102}_{\phantom{1}50}$Sn the admixture of the $T+1$ (in this case $T + 1 = 2$) can be approximated by introducing the Clebsch-Gordan squared coefficient $1/(T+1)=1/2$
\begin{equation}
 \varepsilon_1^2 \approx \frac{1}{2} \bar{\varepsilon}^2.
\end{equation}
The error here is very small. We now apply eq. (\ref{deltacend}). We use the above relation and instead of $\xi \hbar \omega$, we use the energy of the IVM determined in the RPA calculations $\bar{E}_0$. We note that the value of $\xi$ is between 3 to 4. For $V_1$ we take the results of reference \cite{Loc19} to determine the value of $V_1$ by using eq. (\ref{Vsym}). For this purpose, we utilize the RPA results \cite{Loc19} for nuclei that have a neutron excess. For example in the case of $^{42}_{20}$Ca we use the $^{48}_{20}$Ca results. For illustration purposes we show in Table \ref{tab2} the $^{48}_{20}$Ca RPA results. In Table \ref{tab2} we have included also the value of $\delta_C$. Since in $^{48}$Ca the isospin is $T = 4$, the assumption about the equality of the reduced matrix elements for the IVM components with isospins $T+1$, $T$, $T-1$ is not satisfied. Therefore the value of $\delta_C$ is approximate. A rough estimate would assign an uncertainty of $10-15\%$ for the value of $\delta_C$, meaning that for nuclei with a large neutron excess the values of the Coulomb correction are smaller than in the case of super-allowed transitions. The isospin mixing of the $T+1$ states to the ground state is denoted as $\varepsilon_{T+1}^2$ (see reference\cite{Loc19}). When averaging the values obtained with different Skyrme interactions we find the value of $V_1$ for the Ca region to be 90 MeV, for Ni 120 MeV, for Zr 60 MeV, and Sn 90 MeV. Except for Zr, the values of $V_1$ are around 100 MeV. This is the value we used in reference \cite{Loc19}. The Zr region is exceptional, the symmetry energy potential is weaker as noticed a long time ago \cite{Bertsch72}. This point will be mentioned later in the article. The Coulomb potential $V_C$ is computed using the HF calculation. Introducing all mentioned above quantities into eq.~(\ref{deltacend}) we find the total Coulomb corrections $\delta_C$ for the super-allowed beta transitions in $T=1$ nuclei (see Table \ref{tab3}).
\begin{table}[tbh]
\centering
\caption{The Coulomb mixing $\bar{\varepsilon}^2$ (\%) and the IVM determined in the RPA calculation $\bar{E}_0$ (MeV) for $N=Z$ nuclei.}
\label{tab1}
\begin{tabular}{c|r|r||c|r|r}
\hline 
 \multicolumn{3}{c||}{$^{40}$Ca} & \multicolumn{3}{|c}{$^{56}$Ni} \\ \hline
 Skyrme	&	$\bar{\varepsilon}^2$ (\%)	&	$\bar{E_0}$ (MeV)	&	Skyrme	&	$\bar{\varepsilon}^2$ (\%)	&	$\bar{E_0}$ (MeV)	\\ \hline
SIII	&	0.68	&	35.08	&	SIII	&	1.22	&	36.52	\\
SKM*	&	0.78	&	32.51	&	SKM*	&	1.42	&	34.57	\\
SLy4	&	0.77	&	31.13	&	SLy4	&	1.43	&	32.70	\\
BSK17	&	0.70	&	32.79	&	BSK17	&	1.23	&	34.74	\\
SAMi0	&	0.74	&	33.66	&	SAMi0	&	1.36	&	34.57	\\
\hline
 \multicolumn{3}{c||}{$^{80}$Zr}	&	\multicolumn{3}{|c}{$^{100}$Sn} \\ 
 \hline
Skyrme	&	$\bar{\varepsilon}^2$ (\%)	&	$\bar{E_0}$ (MeV)	&	Skyrme	&	$\bar{\varepsilon}^2$ (\%)	&	$\bar{E_0}$ (MeV)	\\ \hline
SIII	&	3.63	&	32.06	&	SIII	&	4.54	&	34.35	\\
SKM*	&	4.07	&	30.18	&	SKM*	&	5.34	&	32.45	\\
SLy4	&	3.96	&	28.93	&	SLy4	&	5.27	&	30.83	\\
BSK17	&	3.72	&	30.21	&	BSK17	&	4.75	&	32.40	\\
SAMi0	&	3.96	&	30.75	&	SAMi0	&	5.15	&	32.58	\\
	\hline
\end{tabular}
\end{table}
\begin{table}[tbh]
\centering
\caption{Results for $^{48}$Ca ($T=4$).} \label{tab2} \vspace{0.25cm}
\begin{tabular}{c|r|r|r|r}
\hline
Skyrme int.	&	$\varepsilon_{T+1}^2$ (\%)	&	$\bar{E_0}$ (MeV)	&	$V_1$ (MeV)	&	$\delta_C$ (\%)	\\
\hline
SIII	&	0.10	&	34.79	&	106.14	&	0.26	\\
SKM*	&	0.12	&	32.54	&	92.17	&	0.28	\\
SLy4	&	0.12	&	30.57	&	97.80	&	0.32	\\
BSK17	&	0.10	&	32.83	&	108.63	&	0.26	\\
SAMi0	&	0.11	&	32.28	&	77.25	&	0.22	\\
\hline
\end{tabular}
\end{table}
\begin{table}[tbh]
\centering
\caption{The Coulomb correction $\delta_C$ (\%) for $T = 1$ nuclei.}
\label{tab3}
\begin{tabular}{c|r|r|r|r}
	&	$^{42}$Ca	&	$^{58}$Ni	&	$^{82}$Zr	&	$^{102}$Sn	\\ \hline
Skyrme	&	$\delta_C$ (\%)	&	$\delta_C$ (\%)	&	$\delta_C$ (\%)	&	$\delta_C$ (\%)	\\
SIII	&	0.40	&	0.54	&	0.70	&	0.98	\\
SKM*	&	0.42	&	0.60	&	0.78	&	1.06	\\
SLy4	&	0.46	&	0.68	&	0.98	&	1.30	\\
BSK17	&	0.44	&	0.62	&	0.78	&	1.14	\\
SAMi0	&	0.32	&	0.52	&	0.68	&	0.98	\\
\hline
\end{tabular}
\end{table}

It is interesting to mention the case of $^{80}$Zr in which isospin mixing was studied experimentally \cite{Corsi11, Ceruti15}. The value for isospin mixing obtained in our work \cite{Loc19} agreed with the experiment.

We should mention that our calculations of $\delta_C$ expresses the global features of this quantity over the periodic table and do not attempt to fit the small fluctuations of this quantity for different nuclei. Our main conclusion is that the $\delta_C$ is smaller by factor $1.5-2$ compared with references \cite{TH, HT20}. The main reason was explained in the Introduction. In our approach, there is no division of the correction into two parts (overlap corrections and the rest). All is taken into account in the single expression eq.~(\ref{deltacend}). Our result for the correction $\delta_C$ is closer to some other computations in reference \cite{Liang09, Rodin13} because in these works some corrections of collective nature of Coulomb strength are taken into consideration. 
\begin{table}[tbh]
\centering
\caption{Results of $\delta_C$ (\%) in various approaches.}\vspace{0.25cm}
\label{tab4}
\begin{tabular}{c|r|r|r}
\hline
	&	$A \approx 40$	&	$A \approx 66$	&	$A \approx 80$	\\ \hline
Hardy-Towner \cite{HT20} &	0.66	&	1.56	&	1.63	\\
Satu\l{}a \textit{et al.} \cite{Satula11} & 0.77 & 0.9 & 1.5-1.6 \\
Rodin \cite{Rodin13}	&	0.43	&	0.99	&	-	\\
Liang \textit{et al.} \cite{Liang09} &	0.33-0.38	&	0.47-0.56	&	1.1-1.2	\\
Auerbach-Loc &	0.40-0.54	&	0.54-0.66	&	0.72-1.12 \\
\hline
    \end{tabular}
\end{table}

\section{Fermi beta transitions, isospin mixing, and the role of the anti-analog state}
So far we have discussed the role of the IVM in inducing isospin impurities into the low-lying nuclear states. The energy of the IVM is high and is distant from the level it admixes. The amount of mixing changes smoothly when going from one nucleus to the next. The IVM involves $2 \hbar \omega$ $ph$ excitations and cannot be properly described in a small space shell-model calculation. When we pass from the parent state $| \pi \rangle$ to the analog nucleus that is the one where one of the neutrons was changed to a proton ($N - 1$ neutrons and $Z + 1$ protons) states with isospin $T - 1$ are allowed. Several states stand out. These are the “configuration” states \cite{Auerbach83, Auerbach72, Bertsch72}. They are composed of the same spatial and spin components as the analog state (denoted as $|A \rangle$)  but are constructed to be orthogonal to the IAS. Of course, they are not eigenstates of the Hamiltonian but are mixed with other $T - 1$ states. So we will treat these states as doorways. The configuration states are expected in general to have relatively large matrix elements with the analog because the Coulomb force produces large monopole contributions \cite{Auerbach83, Auerbach72, Bertsch72}. Among the “configuration” states let us, for the purpose of simplicity, single out one “configuration” state, the anti-analog. In the case that the excess neutrons (or excess protons) occupy only two different orbits, the anti-analog is the only configuration state.

\section{Coulomb mixing of the anti-analog and analog}
Consider a simple parent state in which $n_1$ excess neutrons occupy orbit $j_1(n)$ and $n_2$ neutrons orbit $j_2(n)$. In the parenthesis, we put $n$, or $p$ for neutrons or protons. (In some light nuclei the role of excess neutrons is interchanged with excess protons). Of course, $n_1 + n_2 \equiv N - Z = 2T$ . The parent state is:
\begin{equation}
 | \pi \rangle = \left| j_1^{n_1}(n) j_2^{n_2}(n) \right\rangle
\end{equation}
has isospin $T$. The analog is:
\begin{eqnarray}\label{theanalog}
 |A \rangle = \frac{1}{\sqrt{2T}} \big[\sqrt{n_1} \left|j_1^{n_1 - 1}(n) j_1(p) j_2^{n_2}(n) \right\rangle \nonumber \\ + \sqrt{n_2} \left| j_1^{n_1}(n) j_2^{n_2 - 1}(n) j_2(p) \right\rangle \big]
\end{eqnarray}
has isospin $T$. The anti-analog $|\bar{A} \rangle$ is then:
\begin{eqnarray}\label{theantianalog}
 |\bar{A} \rangle = \frac{1}{\sqrt{2T}} \big[\sqrt{n_2}\left|j_1^{n_1 - 1}(n) j_1(p)j_2^{n_2}(n) \right\rangle \nonumber \\ - \sqrt{n_1} \left| j_1^{n_1}(n) j_2^{n_2 - 1}(n)j_2(p) \right\rangle \big].
\end{eqnarray} 
We consider here parent nuclei with simple configurations: for even-even nuclei, the $n_1$ and $n_2$ are even and in each orbit the excess nucleons are coupled to $J = 0^+$ and in odd-even nuclei $n_1$ is odd and $n_2$ is even. The one-body Coulomb matrix element between the analog and anti-analog is then \cite{Auerbach83, Auerbach72, Bertsch72}: 
\begin{equation}\label{MEVC}
 \langle \bar{A}| V_C |A \rangle = \frac{\sqrt{n_1 n_2}}{2T} \left[ \langle j_1|V_C| j_1\rangle - \langle j_2|V_C|j_2 \rangle \right],
\end{equation}
where $V_C$ is the Coulomb potential. If the excess neutrons occupy orbits belonging to different major shells, this matrix element is sizable. The energy splitting between the analog and anti-analog is often given by the symmetry potential $V_1$:
\begin{equation}\label{Esplit}
 E_{\bar{A}} - E_A = \frac{V_1(N-Z)}{A}.
\end{equation}
The value of $V_1$ is smaller than in the splitting of the IVM and it about 50 MeV (see reference \cite{Bertsch72} and experimental data quoted in this reference).

The coupling between the analog and anti-analog successfully explained \cite{Auerbach83, Auerbach71} isospin forbidden decays in light nuclei \cite{McDonald76, Ikossi76}. As one goes to heavy nuclei along the stability line the number of excess neutrons increases which leads to reductions in the matrix element (\ref{MEVC}) and the increase in the energy splitting (\ref{Esplit}), causing the mixing of $T - 1$ impurities to diminish. The dominant mechanism becomes the mixing with the IVM state. In heavy unstable nuclei with a small neutron excess, the anti-analog mixing mechanism may lead to significant isospin impurities in the analog state. The above example discusses only two orbits, but this mechanism can be easily generalized to more than two orbits.

\section{The anti-analog and the Coulomb correction $\delta_C$} 
We discuss now the contribution of the anti-analog to the Coulomb corrections for Femi beta-decay transitions. Using the definitions in eq.~(\ref{me})
\begin{equation}
 | \Psi_1 \rangle = | \pi \rangle,
\end{equation}
and
\begin{equation}
 | \Psi_2 \rangle = \sqrt{1 - \varepsilon^2} |A \rangle + \varepsilon | \bar{A} \rangle,
\end{equation}
with
\begin{equation}
 \varepsilon = \frac{\langle \bar{A}|V_C|A \rangle}{E_{\bar{A}} - E_A}.
\end{equation}
Note that this is the isospin mixing of the anti-analog into the analog. With these expressions one immediately sees that:
\begin{equation}
 \delta_C = \varepsilon^2.
\end{equation}

\section{Harmonic oscillator estimate}
For a uniform charge distribution with radius $R$ the inner part of the Coulomb potential is
\begin{equation}
 V_C = \frac{1}{2} \frac{Ze^2}{R^3}r^2.
\end{equation}
For $R = 1.2A^{1/3}$ fm one can write the matrix element in equation (\ref{MEVC}) as:
\begin{equation} \label{ME_IAS_GS}
 \langle \bar{A}|V_C|A \rangle \approx 0.35\frac{\sqrt{n_1n_2}}{2T} \frac{Z}{A} \left[\langle j_1|r^2|j_1 \rangle - \langle j_2|r^2|j_2 \rangle \right].
\end{equation}
If $j_1$ and $j_2$ belong to two major shells differing by one node then the difference in the radii square in a harmonic oscillator well becomes:
\begin{equation}\label{Deltar2}
 \Delta(r^2) = \frac{\hbar}{m\omega},
\end{equation}
$m$ is the mass of the nucleon and $\omega$ the oscillator frequency. Taking $\hbar\omega = 41A^{-1/3}$ MeV, we obtain:
\begin{equation} \label{ME_IAS_GS1}
 \langle \bar{A}|V_C|A \rangle = 0.35\frac{\sqrt{n_{1} n_{2}}}{2T} \frac{Z}{A^{2/3}} \rm{MeV}.
\end{equation}
Taking $V_1 \approx 50$ MeV in eq.~(\ref{Esplit}), $N - Z = 2T$
\begin{equation}\label{deltacHO}
 \delta_C = 5.0 \times 10^{-5}Z^2A^{2/3}\frac{n_1n_2}{(2T)^4}.
\end{equation}

\section{Numerical Estimates for the Anti-Analog mixing} 
Using the self-consistent HF potential we computed the difference of the Coulomb matrix elements in eq.~(\ref{MEVC}) for the orbits $2p_{1/2}$ and $1g_{9/2}$ for $^{88}_{38}$Sr. We use the Skyrme HF for the five different forces given previously in the paper \cite{Loc19}. The difference in the Coulomb matrix elements and $\delta_C$ for the above two orbits are shown in Table \ref{tab5}. For the harmonic oscillator the difference in the matrix elements was 0.250 MeV and $\delta_C = 0.14\%$. In nuclei with excess neutrons (protons) occupying two different orbits $j_1$ and $j_2$ but both orbits belonging to the same major harmonic oscillator shell, formula (\ref{Deltar2}) is not applicable. However, it was shown in reference \cite{Auerbach83, Bertsch72, Auerbach71} that due to the different binding energies, and different angular momentum of the two orbits, in a finite well potential, the difference of the two Coulomb matrix elements in eq. (\ref{ME_IAS_GS}) is comparable (within a factor of 2) to the results of the harmonic oscillator. See for example Table 3.2 in reference \cite{Auerbach71}.
\begin{table}[thb]
    \centering
    \caption{$\langle \bar{A}|V_C| A \rangle$ is calculated for $^{88}_{38}$Sr. From the harmonic oscillator estimate $\langle \bar{A}|V_C| A \rangle$ = 0.25 (MeV), and $\delta_C =  0.13\%$.} \vspace{0.25cm}
    \label{tab5}
    \begin{tabular}{c|r|r}
\hline
Skyrme	& $\langle \bar{A}|V_C| A \rangle$ &	$\delta_C$ \\ 
int. & (MeV) & (\%) \\
\hline
SIII	&	0.293	&	0.18	\\
SKM*	&	0.257	&	0.14	\\
SLy4	&	0.281	&	0.17	\\
BSK17	&	0.289	&	0.18	\\
SAMi0	&	0.331	&	0.23	\\
\hline
\end{tabular}
\end{table}

In reference \cite{Bertsch72} in Table 1 are listed a number of Coulomb energy differences for orbits that are within the same major shell or in different major shells. The values are quite similar. One can get an estimate by comparing the relative shifts of states in mirror nuclei. Comparing the low-lying spectra of $^{17}$F and $^{17}$O one finds that the Coulomb energy difference in the parenthesis of eq.~(\ref{MEVC}) for the $s_{1/2}$ and $d_{3/2}$ is about 400 keV. From the spectra of $^{41}$Sc and $^{41}$Ca one finds that the difference in the Coulomb energies for the orbits $p_{3/2}$ and $f_{7/2}$ is about 220 keV and from the spectra of $^{57}$Cu and $^{57}$Ni the difference in Coulomb energies for the $p_{3/2}$ and $f_{5/2}$ is 260 keV. These differences are about half of the Coulomb energy differences found for harmonic oscillator orbits in different major shells. 

It is interesting to mention in this respect that large isospin impurities in the analog have been measured \cite{Melconian11, Bhattacharya08} in the $A = 32$ isobars. In reference \cite{Melconian11} the experiment involved the Fermi transitions within the $T = 1$ isotriplet. The analysis of the experiment indicated a large impurity and a $\delta_C$ correction of 5.3 \%. In the same $A = 32$ nuclei members of the $T = 2$ multiplet were also measured. (The parent state is $^{32}_{18}$Ar$_{14}$ with 4 excess protons). A large isospin impurity of about 1-2\% was found in the analog state \cite{Bhattacharya08}. The shell-model calculation in a restricted space, finds the isospin admixture to be 0.43\% \cite{Signoracci11}. It is remarkable that in these nuclei the primary configuration populated by the excess protons involves two different orbits, the $s_{1/2}$ and $d_{3/2}$, thus allowing for the formation of the anti-analog. If we use equation (\ref{deltacHO}) for the isospin quintet in the $A = 32$, we find $\delta_C = 0.25\%$. The above equation applies to harmonic oscillator orbits belonging to different major shells with $N$ and $N + 1$ nodes, however as already discussed above the difference in the Coulomb matrix elements is affected by the binding energies and angular momentum, and sizable matrix elements between the analog and anti-analog are produced \cite{Auerbach83, Bertsch72, Auerbach71}. Although the mixing with anti-analog might contribute to $\delta_C$ it will not reach the large percentage found in the experiment.

As already remarked proceeding along the stability line to heavier nuclei the number of excess neutrons increases and the isospin admixture caused by the anti-analog decreases. However, presently, (and even more in the future) it will be possible to study, proton-rich, heavy exotic nuclei, with a small neutron (or proton) excess (thus low $T$). In such nuclei the isospin admixtures, as one can see from formula (\ref{deltacHO}), will strongly increase. Choosing nuclei in which the excess protons (neutrons) occupy orbits in different major shells, and have low isospin we can point out some examples (which are not necessarily all feasible for experimental studies). For $T = 3/2$ nuclei with the excess of three nucleons occupying two orbits from different major shells we select $^{17}_{\phantom{1}7}$N$_{10}$ and find from eq.~(\ref{deltacHO}) $\delta_C = 0.04\%$, for $^{44}_{22}$Ti$_{19}$, $\delta_C = 0.7\%$ and for $^{79}_{38}$Sr$_{41}$, $\delta_C = 3.3\%$. For $T = 2$ nuclei one can look at the example of $^{80}_{38}$Sr$_{42}$, here $\delta_C = 2.1\%$. In the examples chosen we selected nuclei in which the excess nucleons occupy two orbits belonging to different harmonic oscillator shells. The trend of fast-growing $\delta_C$ with mass number $A$, for low isospin states and excess neutrons occupying different major shells, is seen in eq.~(\ref{deltacHO}). For $T = 2$ and the mass $A = 40$, $\delta_C = 0.3\%$, for $A = 60$, $\delta_C = 0.9\%$  and for $A = 100$, $\delta_C = 3.9\%$.

\section{$\delta_C$ and the spreading width and energy shifts of the analog}
In the doorway state approximation \cite{Auerbach83, Auerbach72} the spreading width of the IAS is given by
\begin{equation}\label{GammaarrowA}
 \Gamma^{\downarrow}_A = \sum_d\frac{|\langle A|V_C|d\rangle|^2}{|E_A - E_d|^2}\Gamma_d^{\downarrow},
\end{equation} 
where $| d \rangle$ denotes doorway states and $\Gamma_d^{\downarrow}$ their spreading width. 

For the analog, the important doorways are the anti-analog in lighter nuclei and the IVM in the heavier. Here we limit ourselves to the anti-analog $| A \rangle$. Eq. (\ref{GammaarrowA}) can be written as
\begin{equation}
 \Gamma^\downarrow_A = \delta_C\Gamma^\downarrow_{\bar{A}}.
\end{equation}
The spreading width of the anti-analog $\Gamma^\downarrow_{\bar{A}}$ is due to the strong interaction and therefore is of the order of a single-particle spreading width, thus several MeV.

The practicality of the above equation is quite limited. The total width of the analog state is in general composed of the escape width \cite{Auerbach83, Auerbach72, Auerbach71} and spreading width. It is usually difficult to separate the two. And then the spreading width of the analog gets a contribution from the IVM and anti-analog and again it is not easy to separate the two. As we just mentioned above, the spreading width of the anti-analog is not well known. One can get only a rough idea about the contribution of the anti-analog to the width of the analog. For example, if we use a 3 MeV spreading width for the anti-analog in $^{88}$Sr and the estimated value for $\delta_C$, we conclude that the contribution of the anti-analog to the spreading width of the analog is only a few keV. However, it is worth noticing that in some medium mass nuclei, in low-isospin exotic nuclei, the mixing with the anti-analog can produce relatively large spreading widths of the analog, of the order of a few tens of keV. For example, in $^{79}_{38}$Sr$_{41}$ the contribution of the anti-analog to the spreading width of the analog is of the order of 100 keV. In experiments, one would observe broadened analog resonances. The mixing discussed here affects also the energies of the IAS. This mixing might produce shifts of the order:
\begin{equation}\label{deltaEepsilon2}
    \Delta E = \delta_C(E_{\bar{A}} - E_A).
\end{equation}
The values of $\Delta E$ are typical of the order of several tens of keV. For example in $^{79}_{38}$Sr$_{41}$ this equals 70 keV. The shifts may vary for different states in the analog nucleus and the spectrum maybe somewhat distorted compared to the parent nucleus. For example, the three excess neutrons may occupy states with $J = 5/2, J=1/2$ of the kind $p_{1/2}^2f_{5/2}$, $p_{1/2}f^2_{5/2}$ while another configuration with quantum number $J = 5/2$ will be $f^3_{5/2}$. The first two states will have anti-analogs, the third will not have and therefore the first two states will be shifted according to eq. (\ref{deltaEepsilon2}) while the third will not.

A short account of this work was discussed at a conference in 2014 in section 4 of reference \cite{Auerbach14}.

\section{Conclusions and Outlooks}
First we summarize the first part, the
Coulomb corrections to super-allowed beta decay.
We stress here again, the main purpose of the first part of our work is to calculate the total correction to $\delta_C$ using the Coulomb interaction unchanged and taking into account the collective effects of the particle-hole space in the isovector channel. It is not new that such collectivity causes the states to shift to higher energies. The discovery of the giant electric dipole was found at excitation energy of $2\hbar\omega$ instead of $1\hbar\omega$  expected from non-collective particle-hole states. The same also holds for the isovector $J = 0^+$  channel. This would cause a reduction in isospin mixing in the ground states of even-even nuclei. In turn, this leads to reduced values of the $\delta_C$. The purpose of the present paper (and the one in reference \cite{Auerbach09}) was to include this effect and not to find all relevant corrections to the superallowed beta decays. We did not intend to calculate the $V_{ud}$ matrix element in the CKM matrix. The other approaches \cite{TH, Liang09, Satula11, Rodin13, Xayavong18, Ormand95} do not include the effect of the collective shift of the Coulomb strength. So far none of these works explain how one can avoid it, and how other aspects of their theory are able to compensate for this nuclear structure effect. In reference \cite{Auerbach09} we suggested a simple model to account of this collective aspect of theory and how it affects the amount of isospin mixing. In the present paper, we use the results obtained in reference \cite{Loc19}. The work in \cite{Loc19} is a fully microscopic, detailed method to find the isospin admixture in the ground states of even-even nuclei. This is probably, at present, one of the best calculations of isospin mixing in the ground states of even-even nuclei. The calculation is employing many versions of the Skyrme interaction. We were able to separate the three isospin components ($T-1, T$, and $T+1$) of the isovector excitations, and were able to determine their energy splitting. From there we found the values of the symmetry potential $V_1$, and the excitation energies of the IVM state.  Indeed the energies turn out to be at $3\hbar\omega$ and not at $2\hbar\omega$. The new values for these quantities were used in the present paper. The values obtained for the $\delta_C$ in the present paper are about a factor of $2$ smaller than the ones found by Hardy and Towner \cite{TH, HT20} (see Table \ref{tab4}).

Concerning the Conserved Vector Current (CVC) hypothesis test, which really means, in the present context, that the $\mathcal{F}t$ values should be constant with $Z$ and $A$. This would be the case when the isospin symmetry is fully conserved. If all corrections are introduced to the measured values then this will be the case. The $\delta_C$ is only one among a few other corrections. As explained above we only calculate the $\delta_C$. Hardy and Towner when they calculate this correction for the various nuclei they adjust several parameters in their theory for each nucleus separately. Their model is semi-phenomenological and the strength of the two-body Coulomb interaction is adjusted to fit the experimental Isobaric Multiplet Mass Equation (IMME) for each nucleus under consideration. Also, a charge-dependent nuclear interaction is incorporated by a 2\% increase in all the $T = 1$ proton-neutron matrix elements in Hardy and Towner work \cite{TH}. It could affect isospin mixing in certain cases, when the levels that mix are close in energy, and the two-body matrix element may affect the mixing. This could happen to some close-spaced excited levels, or in odd-odd nuclei. In our approach, we do not calculate the Coulomb correction for each nucleus and do not adjust the interaction in each case. No surprise that we do not consider the CVC theorem.

It is of interest to note the following. Recently Hardy and Towner published a detailed survey of the Superallowed transitions \cite{HT20}. The main and ultimate purpose of Hardy and Towner is to use the Superalowed beta transitions in order to determine with great precision the value of the $V_{ud}$ matrix element in the CKM matrix and to assess whether unitarity is fulfilled or violated. In our work, the purpose is just to consider one aspect of the theory namely how the collectivity of the state that causes isospin mixing affects $\delta_C$. The aims of our work and the one of Hardy and Towner are quite different and one should not compare the results of these two approaches. There is extensive work in which other corrections are considered \cite{TH, Seng18, Seng19}. However, if we adopt the various values of the corrections listed in Ref.~\cite{HT20}, except $\delta_C$ of course, and insert our values of this parameter we obtain for $A = 60$ the $V_{ud}^2$ to be smaller by $1\%$ than what was obtained in Hardy and Towner \cite{HT20}. This is a very rough estimate but it indicates that in spite of the large difference in the values of $\delta_C$ in the two approaches the change in $V_{ud}^2$ is less than 1\%. So at the level of low precision requirement, the value of $V_{ud}^2$ is not very sensitive to the value of $\delta_C$. This is not surprising, the percentage of all Coulomb corrections is very small.

In our approach, the calculation of isospin mixing is detailed and advanced. It can be applied to any even-even nucleus with any isospin in the ground state. Applying the results of the new calculations of isospin mixing to obtain $\delta_C$ in nuclei with $T>1$ requires additional improvement. We plan to improve this by calculating explicitly the reduced matrix elements for the $T-1, T$, and $T+1$ for the IVM state. The overestimate of the amount of isospin mixing described by non-collective single-particle models was already noted a long time ago (see for example \cite{Auerbach83, BMVolI, Auerbach72, Yeverechyahu76, Bertsch72}. For example, isospin mixing determines the spreading width of the Isobaric Analog Resonance (see reference \cite{Auerbach83, Auerbach72}). When using the single-particle model and one-particle optical potential the spreading widths for the Isobaric Analog Resonance turn out to by factors 5 (or more) larger than the experimental ones. It was noted that introducing some correlations among various particle-hole states one can reduce this very large discrepancy \cite{Auerbach83, Auerbach72}, but this still remains a problem until today.

Now we summarize the second part, the influence of the anti-analog state on isospin mixing in the isobaric analog state and the correction to the beta-decay Fermi transition.
Recently a paper was published \cite{Hoff20} in which a considerable deviation from isospin symmetry was observed for a $T=3/2$ isospin multiplet in the Sr region. Following that paper, an article was published \cite{Lenzi20} in which an attempt was made to explain the results in \cite{Hoff20} using a shell-model approach.

The anti-analog state (or any other configuration states) are not eigenstates of the full Hamiltonian and are fragmented by the strong force. Here we treated the anti-analog as a doorway, assigning it an average energy position and placing there its unperturbed configuration. This of course is an approximation. The admixtures found in the restricted shell-model are small, not exceeding 0.5\%, and usually, of the order of 0.1\% \cite{Signoracci11}. It is not clear whether these calculations include fully the anti-analog mixing caused by the one-body Coulomb field. Our approach here is more transparent and does not require complicated computations. It would be nice to know whether indeed in shell-model calculations the large contribution to the isospin impurities comes from components that make up the anti-analog.

Our calculations depend on the values of several parameters which are not well determined. The value of the energy separation between the analog and anti-analog is uncertain. The fragmentation of the anti-analog strength will affect the outcome, and of course, the structure of the parent state is important. Even when the basic configuration of the parent state does not involve two different orbits, configuration mixing will bring in some higher orbits, which would validate the anti-analog mechanism. Although this is a second-order effect in the evaluation of the isospin impurity, in some cases configuration mixing is substantial and the admixtures of configurations with higher orbits could be large, thus leading to sizeable anti-analog components. One should also mention the iso-multiplets of excited states. In this case, one can often find situations in which higher orbits (from different major shells) compose the excited states. The two-body part of the Coulomb force is neglected in our approach. A shell model calculation takes into account the two-body part of the Coulomb interaction.

The isospin mixing in the analog due to the discussed mechanism has a particular dependence on the mass $A$, the charge $Z$, and the excess neutron number $(N - Z)$. It is conceivable that in studies of exotic nuclei one can choose favorable cases where isospin mixing is large and learn more about this subject.

\section*{Acknowledgements}
We thank Chien-Yeah Seng and Dai-Nam Le for the discussions.

\section*{Declaration of competing interest}
The authors declare that they have no known competing financial interests or personal relationships that could have appeared to influence the work reported in this paper.

\bibliography{mybibfile}

\end{document}